\documentclass{aa}
\usepackage{graphicx}
\usepackage{txfonts}
\usepackage[authoryear]{natbib}
\begin{document}
%
%   \title{Mid-infrared meteoritic nanodiamond emission in 3C\,298: a negative result}
   \title{Quasar 3C298:  a test-case for meteoritic \\nanodiamond 3.5\,\mic\ emission}

%   \subtitle{}

% CITATIONS
%   Verba que praevisam rem non invita sequentur. (Horatius)
%   The words follow effortless the things that were previously
%   thought. (Horace)
%  LUC: encouraging words when writing a paper!

   \author{J. A. de Diego
   \inst{1}
          \and
          L. Binette
          \inst{1}
          \and
         P. Ogle
         \inst{2}
        \and
          A.C. Andersen
          \inst{3}
          \and
          S. Haro-Corzo
          \inst{1,4}
          \and
          M. Wold
          \inst{5,6}
          }
%          C. Ptolemy\inst{2}\fnmsep\thanks{Just to show the usage
%          of the elements in the author field}

 %  \offprints{J. A. de Diego}

   \institute{Instituto de Astronom\'{\i}a,
             Universidad Nacional Aut\'{o}noma de M\'{e}xico,
             A.P. 70-264,
             04510 M\'{e}xico D.F.,
             Mexico
%             \\ \email{jdo@astroscu.unam.mx; LBinette@astroscu.unam.mx}
         \and
             Spitzer Science Center, MS 220-6, California Institute of Technology,
             Pasadena, CA 91125, USA
         \and
             Dark Cosmology Centre, Juliane Maries Vej 30, DK-2100 Copenhagen, Denmark
        \and
           Present address: Instituto de Ciencias Nucleares, Universidad Nacional Aut\'{o}noma de M\'{e}xico,
            A.P. 70-543, 04510 M\'{e}xico D.F., Mexico
         \and
             European Southern Observatory, Karl-Schwarzschild str. 2, 85748
             Garching bei M\"unchen, Germany
         \and
            Present address: Institute of Theoretical Astrophysics,
            University of Oslo, P.O. Box 1029, Blindern, 0315 Oslo, Norway
              }

   \date{Received January 11th; revised February 8th; accepted 1st March}

%________________________________________________________________
%
% LIST OF COMMANDS
%
% Ions
  \newcommand{\halpha}{\ion{H}{$\alpha$}}
  \newcommand{\hbeta}{\ion{H}{$\beta$}}
  \newcommand{\hgamma}{\ion{H}{$\gamma$}}
  \newcommand{\oi}{[\ion{O}{i}]}
  \newcommand{\oiii}{[\ion{O}{iii}]}
  \newcommand{\nii}{[\ion{N}{ii}]}
  \newcommand{\sii}{[\ion{S}{ii}]}
  \newcommand{\hi}{\ion{H}{i}}
  \newcommand{\hii}{\ion{H}{ii}}
  \newcommand{\feii}{\ion{Fe}{ii}}
  \newcommand{\fevii}{[\ion{Fe}{vii}]}
  \newcommand{\fex}{[\ion{Fe}{x}]}
  \newcommand{\hei}{\ion{He}{i}}
  \newcommand{\heii}{\ion{He}{ii}}
%
% Other commands
  \newcommand{\mic}{\hbox{$\mu{\rm m}$}}  %micrometer unit
  \newcommand{\kms}{km~s$^{-1}$}    % Velocity km/s
  \newcommand{\lb}[1][a]{$^{#1}$}   % Label for tables: default a
  \newcommand{\bv}{\textit{B-V}}    % Color B-V
  \newcommand{\cms}{\hbox{${\rm cm^{-2}}$}}  %column units
  \newcommand{\gcc}{\hbox{${\rm g\,cm^{-3}}$}}  %density units
\newcommand{\coms}[1]{{}{\bf #1}}  %% \coms{main comments}
\newcommand{\comc}[1]{{$\clubsuit$}{\it #1}}  %% \comc{style comments}

%\renewcommand{\coms}[1]{#1}  %% hide marking
%\renewcommand{\comc}[1]{#1}  %% hide marking
%________________________________________________________________

% \abstract{}{}{}{}{}
% 5 {} token are mandatory

  \abstract
   {}
   {We calculate the dust emission expected at 3.43 and
   3.53\,\mic\ if meteoritic (i.e. hydrogenated) nanodiamonds are
   responsible for most of the far-UV break observed in quasars.}
   {We integrate the UV flux that hydrogenated nanodiamonds must
   absorb to reproduce the far-UV break. Based on laboratory spectra
   of H-terminated diamond surfaces, we analyse the radiative energy
   budget and derive theoretically the IR emission profiles expected
   for possible C$-$H surface stretch modes of the diamonds.}
   {Using as test case a spectrum of 3C298 provided by the Spitzer
   Observatory, we do not find evidence of these emission bands.}
   {While diamonds without surface adsorbates remain a viable
   candidate for explaining the far-UV break observed in quasars,
   hydrogenated nanodiamonds appear to be ruled out, as they would give
   rise to IR emission bands, which have not been observed so far.}


   \titlerunning{Meteoritic nanodiamond emission in quasars}

   \maketitle

%
%________________________________________________________________

\section{Introduction}

% PROBLEM

The spectral energy distribution (SED) of  quasars is composed of
emission lines superimposed on various continuum emission
components. The optical to near-infrared domain is reasonably well
reproduced by a power law. The far-infrared and  ultraviolet
spectral regions are both characterised by a broad continuum excess
with respect to the optical-IR power law. The  UV flux excess is
referred to as the big blue bump (BBB). According to general belief,
it corresponds to the thermal signature from a hot accretion disk
orbiting a supermassive black hole. The extension in the extreme UV
of the BBB is expected to provide the ionising flux that powers most
of the emission lines.
% UV break

% CAUSES
%   Haec neque affirmare, neque refellere operae pretium est
%   famae rerum standum est. (Titus Livio)
%   It is not worth either confirm or deny these things, let
%   us rely on what is told.
% HI absorption
%Other explanations using \hi\ have been less successful. Intrinsic
%\hi\ absorption would produce a sharp absorption edge at 912\,\AA\
%as well as a pronounced Ly$\alpha$ absorption line, which are not
%observed in quasar spectra. Recently, \citet{bin03} studied \hi\
%absorption by the intergalactic medium, but they had to reject this
%explanation because their model predicted a jump in flux blueward of
%1216\AA\ in the observer-frame that is not seen. The FUV break
%spectral signature has also been approached by comptonized accretion
%disk model \citep{zhe97} and by state of the art 'naked' accretion
%disk models \citep{hub00}. Neither of these disk models can
%reproduce the flux upturn at higher energies, hence the softness
%problem remains unresolved.

A serious problem with this picture, however, is that the BBB
observed in quasars appears to be too soft to account for the high
excitation lines from the broad emission line region (BELR)
\citep[see][]{kor97}. In effect, a marked continuum drop or
steepening takes place shortward of $\simeq 1100$\,\AA, which we
hereafter term the far-UV break. This break is observed in the
composite quasar SED derived by \citet{tel02}, as well as in
individual spectra of quasars with $0.7 \la z \la 2$ \citep[
hereafter B05]{bin05}. A possibility might be that this break is
more akin to a localised continuum trough, followed by a marked
recovery in the extreme UV, which is the energy region responsible
for the high-excitation emission lines. Various mechanisms that
could generate such a trough are summarised by \citet{bin07a}.
State-of-the-art calculations of the SED for standard
geometrically-thin, optically-thick accretion disks do not reproduce
the observed break satisfactorily, but these models assume a
stationary disk with a vertical structure supported only by gas and
radiation pressure. To our knowledge, no detailed SED calculations
have been carried out assuming a non-stationary disk or with an
accelerating wind. Absorption by intergalactic \hi\ and by
intergalactic dust has been discarded by \citet{bin03} and B05,
respectively. \citet{eas83} proposed that \hi\ absorption by local
clouds that are accelerated up to 0.8$c$ could generate a steepening
of the transmitted continuum. To reproduce modern data, this model
would need to be fine-tuned and extended to the extreme UV where the
flux recovery is expected. More recently, B05 propose that the break
could be the result of absorption by crystalline dust grains {\it
local} to the quasars. They successfully reproduced the position and
detailed shape of the UV break in 50 quasars, out of a total sample
of 61 objects from HST-FOS archives whose spectra extended down to
at least 900\,\AA\ (rest-frame). Their model required two flavours
of crystallites: nanodiamonds {\it without} surface adsorbates and
nanodiamonds {\it with} surface adsorbates as found in most
measurements of meteoritic nanodiamonds from primitive carbonaceous
chondrite meteorites; see \citet[ hereafter MA04]{mutschke04} and
references therein, and also \citet[ hereafter JH04]{Jones_etal04}.
If the far-UV break {is} due to dust absorption, we expect to
observe the infrared (IR) re-emission corresponding to the absorbed
energy. This is not a straightforward test in the case of pure cubic
diamonds, since only multi-phonon modes are active in the IR
\citep[JH04;][]{braatz00a, edwards85}. However, impurities or
structural defects can enable one-phonon modes to become active
\citep{anders98, JonesHendecourt00}.

\citet{guillois99} and \citet[][ hereafter VK02]{ker02} make a
convincing case for the detection of hydrogenated nanodiamonds
within circumstellar media for which the C$-$H stretch emission
bands at 3.43\,\mic\ and 3.53\,\mic\ were observed\footnote{Pure
diamonds are extremely {\it inefficient} emitters compared to the
one-phonon mode provided by vibrational modes of carbon bonds with H
atoms, which populate the surface of meteoritic nanodiamonds.
Therefore, considering only one-phonon modes for the cooling of the
grains is considered an acceptable approximation.}. As laboratory
measurements of meteoritic nanodiamonds have shown them to be
surface H-terminated \citep[ MA04]{lewis89, hill97, anders98,
braatz00a}, we can test their presence by using a method similar to
VK02. We adopt the spectra of the quasar 3C298 as an example
test-case, since UV and mid-IR coverage are both available from the
HST and Spitzer archives.

% Dust emission

%The nanodiamods considered consist either of terrestrial cubic
%diamonds or with surface impurities as found in carbonaceous chondrite
%meteorites, such as Allende. The flux upturn taking place below
%650\,\AA\ near allow the intrinsic SED to be much harder than
%indicated by extrapolating of the flux near the FUV break. Such an
%upturn was identified in 4 quasar spectra of the \citet{tel02}
%sample. From a spectrum that combines various archives, \citet{bin06}
%reported a FUV rise in Ton\,34.

%__________________________________________________________________

\section{Calculations of IR emission from hydrogenated diamonds}

% Quid fecisti? -- What have you done?

% EMISSION(IR) = ABSORPTION(UV)

% Van Kerckhoven et al. 2002 - Section 7

Our first step consists in estimating the far-UV energy absorbed by
hydrogenated (meteoritic) nanodiamonds, following the prescription
of B05. The second step consists in defining a geometry for the dust
and computing the IR emission profile using the same method as VK02.
The derived profiles are finally compared to the mid-IR observations
of 3C298.

We assume the quasar continuum source to be isotropic and surrounded
by a dust screen that lies some distance away from it. For the sake
of simplicity, the screen can be viewed as a spherical shell that
covers a fraction $\Omega$ of the source's sky. The continuum source
presumably consists of an accretion disk. Although the dust emits
isotropically,  the far-side dust screen is not directly observable,
due to the opaqueness of the (intervening) accretion disk,
%(no pass-through of back shell emission)
which accounts for the factor $1/2$ in Eq.\,\ref{eq:fir} below.
% POWER LAW
The intrinsic SED over the whole UV domain is modelled as a power
law of index $\alpha$, with $F_{\nu} \propto \nu^{+\alpha}$ and
$F_{\lambda} \propto \lambda^{-(2 + \alpha)}$. To derive a more
general formulation of the dust absorption trough, we use the
concept of equivalent width $EW_{\rm A1}$, which we define relative
to the continuum flux at 1350\,\AA, longward of the break, which is
a spectral region relatively free of lines (e.g. Fig.\,5 of B05).
The total energy absorbed by the  dust and expected to be re-emitted
in the IR is given by
\begin{equation}\label{eq:fir}
F_{em}^{IR}=\frac{\Omega}{2} \, F_{1350}^{obs}\, EW_{\rm A1} \; ,
\end{equation}
where $F_{1350}^{obs}$ is the \emph{observed} flux at 1350\,\AA\
(rest-frame) for the quasar of interest. The value of $EW_{\rm A1}$
is given by the integral
% EQUIVALENT WIDTH
\begin{equation}\label{eq:ew}
{EW_{\rm A1}} = \frac{\int^{^{1300\,\AA}}_{_{300\,\AA}} \! \! \! \!
(1-T_{\lambda}) \; F_{\lambda} \; d\lambda}{F_{1350\,\AA}} \;,
\end{equation}
where $T_{\lambda}= \exp{(-\tau_{\lambda}})$ is the transmittance of
the dust in terms of its opacity $\tau_{\lambda}$, and $F_{\lambda}$
the UV flux impinging on the dust screen. The opacity can be
formulated in terms of the screen column $N_{\rm H}$ times the
extinction curve cross-section normalised to H. In the model of B05,
the dust typically consists of 40\% of nanodiamonds without surface
adsorbates and 60\% of hydrogenated nanodiamonds. Hence, the opacity
due to hydrogenated nanodiamonds from meteorites is $0.6 N_{\rm H}
\, \sigma_{\lambda}^{\rm A1}$, where the latter cross-section
corresponds to the extinction curve A1 calculated by B05 for
hydrogenated nanodiamonds, assuming the small grain size regime. To
describe the SED, we adopt a spectral index of $\alpha = -0.45$,
which is essentially the mean  value encountered by B05 among their
39 class\,A quasars\footnote{The near-UV coverage provided by the
HST-FOS spectrum of 3C298 was insufficient to constrain  $\alpha$
effectively.}. B05 find that the H column that fitted the UV break
of their class\,A quasars was $1.02\pm{0.29} \times 10^{20}\,$\cms.
We obtain a satisfactory fit of the far-UV break in 3C298 by
assuming a column of $1.2\times 10^{20}$\,\cms, the value assumed
hereafter. Integrating Eq.\,\ref{eq:ew}, we derive a value of
$EW_{\rm A1}$ of 930\,\AA, which we consider representative of the
majority of the class \,A spectra studied by B05.

To get a handle on the dust covering factor $\Omega$, we define, on
the one hand,  the maximalist case where the source is covered in
all directions by the dust screen, and on the second, a minimalist
case, in which the dust only covers the source's sky as defined by
the ionising bicone. The assumed bicone aperture is 45\degr. The two
cases can be summarised as follows:
\begin{eqnarray}
  \textrm{Case A:} \quad \Omega &=& 1, \label{eq:a}\\
  \textrm{Case B:} \quad \Omega &=& 1 - \cos
  (45\degr/2) = 0.076 . \label{eq:b}
\end{eqnarray}
With case\,B, we suppose that the UV radiation that emerges outside
the ionising bicone is absorbed by the opaque torus rather than by
crystallite dust. Since the UV break is ubiquitous in quasars, we
consider the minimalist case\,B as a plausibly strict minimum for
$\,\Omega$. The difference between the two cases amounts to an order
of magnitude and represents the level of uncertainty regarding
geometrical considerations.

\section{The quasar 3C\,298 as test-case}

% WHY 3C 298
%Faint Object Spectrograph (FOS) observations with HST have been
%carried out for this quasar.

The well-known radio-loud quasar 3C\,298 has a redshift of $z=1.436$
that allows the Spitzer IR Observatory to cover the relevant
3.5\,\mic\ region (rest-frame), where emission from C$-$H stretch
modes is expected to occur. We retrieved the S14.0 IRS pipeline data
from the Spitzer archive. The exposure times were 240\,s with the
Short-Low\,2 (SL2+3) slit and 120\,s with the Short-Low 1 (SL1) and
Long-Low\,2 (LL2) slits. The slit widths are 3\farcs6, 3\farcs7, and
10\farcs6, respectively. The spectral resolving power is 86 at
5.2\,\mic\ (2.1\,\mic, rest) in SL2, 71 at 8.5\,\mic\
(3.5\,\mic,rest) in SL1 and SL3, 86 at 14.6\,\mic\ (6\,\mic, rest)
in LL2, and it increases linearly with wavelength in each order. Two
cycles were averaged, then nods were subtracted to remove the
background emission. Spectra were extracted in standard apertures
(SL2: 7\farcs2 at 6\,\mic, SL3: 7\farcs2 at 8\,\mic,  SL1: 14\farcs4
at 12\,\mic, and LL2: 21\farcs7 at 16\,\mic). All orders and nods
are well-matched in flux, indicating accurate pointing on the slits
and no significant extended emission.

The resulting spectrum is shown in Fig.\,\ref{fig:3C298a}. Within
the region of interest, 3--4\,\mic, it agrees within 3\% with the
published (smoothed)  version of the same spectrum by
\citet{haas05}\footnote{The flux scaling in their Fig.\,1 should
read $10^{14}$ rather than the misprinted value of $10^{12}$
\citetext{M.\ Haas,\ priv.\ comm.}.}. However, we note that our IRS
fluxes are lower by a factor 2.5 compared to those quoted in the
NASA/IPAC Database (NED), which consisted of ISO and Spitzer MIPS
photometric measurements. Variability of the underlying non-thermal
component, rather than of the dust, is the most likely explanation
\citep{cleary07}.

% QUANTITIES
% Flux(uv)

The FOS-archived spectrum for this object indicates a flux value of
$F_{1350}^{obs} = 1.0 \times 10^{-15}\, \textrm{erg\, s}^{-1}\,
\textrm{cm}^{-2} \AA^{-1}$ (corrected for Galactic reddening). Using
Eqs.\,\ref{eq:fir} and \ref{eq:ew}, we derived integrated flux
values of $F_{em}^{IR}= 4.7 \times 10^{-20}$ and $0.36 \times
10^{-20}\, \textrm{W} \, \textrm{cm}^{-2}$ for the $\Omega$ cases\,A
and B, respectively. {}That much energy should appear as two IR
bands at 3.43 and 3.53\,\mic.

% Flux(ir)

\subsection{The 3.43 and 3.53 IR bands from HD\,97408 data}\label{subvk}
\begin{figure}
   \centering
   \includegraphics[width=\linewidth,clip]{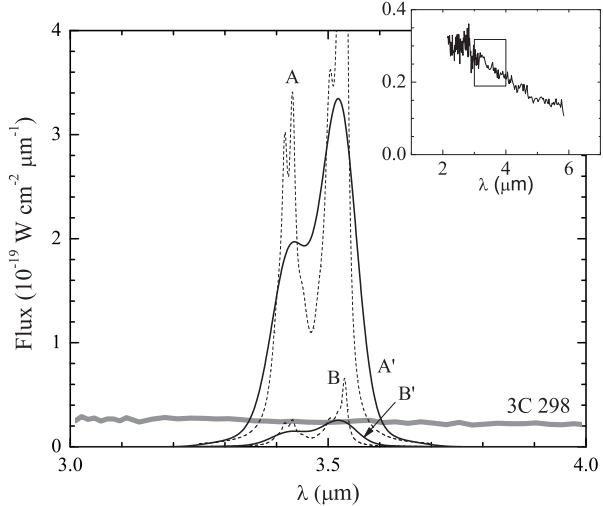}
\caption{Spectrum of 3C298 and predicted emission bands {based on
the empirical work of VK02.} The thick gray line shows the Spitzer
mid-IR spectrum of 3C\,298 around the 3.5\,\mic\ region (rest-frame
wavelengths). A typical errorbar (with $\sigma \sim
0.02\times10^{-19}\,\textrm{W} \, \textrm{cm}^{-2} \, \mic^{-1}$) in
the interval 3.2--3.8\,\mic\ is much smaller than the gray line's
thickness. The inset reveals the Spitzer spectrum over a wider
spectral domain. The predicted emission from hydrogenated
nanodiamonds, assuming cases\,A and B, is shown by dotted lines.
Continuous lines (labelled A$^\prime$ and B$^\prime$) show the
convolution of the VK02 profiles with the spectral resolution of the
IRS spectrograph ($R \simeq 60$) (see Sect.\,\ref{subvk}).
%They are meant to be compared with the 3C\,298 spectrum.
%%%%%%%%%%and a velocity field of 4000\,\kms\ (FWHM).
}
\label{fig:3C298a}%
\end{figure}

To model the nanodiamond emission, we adopted the 8-components fit
of VK02 of the two bands\footnote{VK02 provide a comprehensive and
self-consistent model of the IR bands, in which the stellar UV
radiation heats up the dust. The profile of the strongest component
at 3.53\,\mic\ allowed these authors to infer a nanodiamond
temperature of $\sim 900$\,K and a grain size range of
10--100\,\AA.} that were observed in the ISO spectrum of the stellar
disk of Herbig Ae/Be star HD\,97408. The method used by VK02 was to
decompose the observed 3 micron region using Lorentzian profiles and
to determine the peak frequency, width, and relative strength of the
profiles by a least-square fitting routine. The other two stars
studied by these authors show similar features, but with a lower
S/N. The profile was normalised to reflect the $F_{em}^{IR}$ values
expected in 3C298 and is shown in  Fig.\,\ref{fig:3C298a} for
cases\,A and B.
%The nanodiamond emission calculated for cases\,A and B are overlaid
%using thin continuous lines.
%Note that the flux units reported by \citeauthor{haas05} had to be divided
%by a factor 100 to correct for an error of labels used in the published graph
%\citetext{M.\ Haas,\ priv.\ comm.}.
The continuous lines show the result of convolving the nanodiamond
emission profiles with the spectral resolution characterising the
data on 3C298 ($\lambda / \Delta \lambda \simeq 60$).
% and a dust velocity field of 4000\,\kms\ (FWHM).

% One column figure
%______________________________________________ IR spectrum 3C 298
%

% sub-DISCUSSION

As shown in Fig.\,\ref{fig:3C298a}, {hydrogenated} nanodiamond
emission should be detected clearly, if present. In the case\,A
scenario, the prediction is of very strong emission bands (thin line
A$^\prime$ peaks at $\simeq 14$ times above the quasar continuum
height) that should swamp the IR spectrum at wavelengths near
3.5\,\mic, which is not observed. For case\,B, where the emission
only originates  within the near-side ionising cone of the quasar,
the nanodiamond emission is much less dramatic but still comparable
to the height of the continuum. The lack of any dust signature on
the IR spectrum of 3C\,298 appears to rule out that hydrogenated
nanodiamonds similar to the meteoritic type are present in the
amount predicted by the UV absorption model of B05.

\subsection{The 3.43 and 3.53\,\mic\ bands from the laboratory}\label{sublab}

%{Laboratory measurements of the 3.43 and 3.53\,\mic\ absorption
%bands due to hydrogenated nanodiamonds from the Orgueil meteorite
%were performed by JH04. They compared these with the bands produced
%by larger synthetic diamonds appropriately hydrogenated and found
%that the spectra observed in HD\,97408 and Elias\,I matched better
%the profiles resulting from the larger hydrogenated diamonds ($>
%50$\,nm), while the Orgueil meteoritic nanodiamonds showed a profile
%lacking a strong dip in between the two bands.

\begin{figure}
   \centering
   \includegraphics[width=\linewidth,clip]{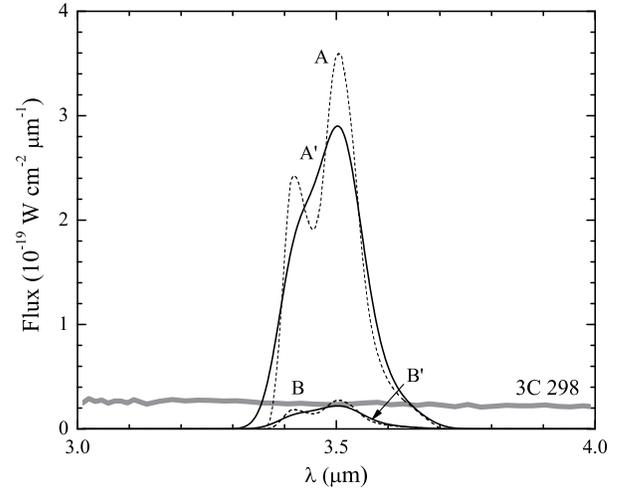}
\caption{Spectrum of 3C298 and predicted emission bands {based on
absorption profiles measured from nanodiamonds of the Orgueil
meteorite, as measured by JH04. The notation is the same as in
Fig.\,\ref{fig:3C298a}.
%(see Sect.\,\ref{sublab})
}}
\label{fig:3C298b}%
\end{figure}

Independent laboratory measurements of nanodiamonds from distinct
meteorites show different peak positions and strengths for the
features around 3.43 and 3.53\,\mic\ \citep[see Table\,1
in][]{braatz00a}. It is not clear whether these variations disclose
a real physical difference between the diamond populations from
different carbonaceous chondrites, or if they are the result of the
laboratory procedures. MA04 used 100\,g of the Allende meteorite,
while JH04 used 1\,g of the Orgueil meteorite for their otherwise
almost identical laboratory measurements. The data by MA04 show
similar features  to those observed by VK02, while the data from
JH04 do not present the same features. JH04 compared the Orgueil
data with the bands produced by larger hydrogenated synthetic
diamonds and, in contrast to VK02, conclude that the spectra
observed from HD\,97408 and Elias\,I are better matched by the
larger diamonds' profiles. Supposing that JH04 is right, instead of
using the empirical profiles from HD\,97408, we should then consider
using the profiles measured by JH04 (their Fig.\,1) corresponding to
the nanodiamonds of the Orgueil meteorite. In this way, we would
ensure that the grain size regime remains consistent with the B05
model, which relies on a dust model containing meteoritic
nanodiamonds of radii $< 0.6$\,nm. In Fig.\,\ref{fig:3C298b} we
present the expected IR bands, assuming the JH04 profiles inferred
from the nanodiamonds of the Orgueil meteorite. The areas under the
profiles and the notation are the same as in Fig.\,\ref{fig:3C298a}.
As can be seen by comparing both figures, the conclusions reached do
not depend on whether we follow the interpretation of VK02 or of
JH04.

%%%concluding DISCUSSION

% One column figure
%______________________________________________ IR spectrum 3C 298

\section{Discussion}

% Si finis bonus est, totum bonum est
% If it finishes OK, everything is right
% but it does not with nanodiamonds! :(

No emission bands are observed in 3C298 at 3.43 and 3.53\,\mic\
(rest-frame), which we could ascribe to hydrogenated diamond C$-$H
stretches, whether we assume the VK02 or the JH04 profiles. Before
concluding that meteoritic (i.e. hydrogenated) nanodiamonds are
ruled out in quasars, we must contend with the following issue. We
have assumed that {\it all} the energy absorbed in the UV was to
reappear entirely within the 3.43 and 3.53\,\mic\ bands.
\citet{JonesHendecourt00} differ on this point, as they propose that
the nanodiamonds account for many of the PAH features as well.
%which are observed in the IR spectra of Galactic objects.
However, VK02 argue against this view. On the other hand, even if we
considered the whole of the 7.7\,\mic\ feature and the 20\,\mic\
plateau observed in HD\,97408 to be the results of nanodiamond
emission, this would still leave 30\% of the energy to be emitted at
3.43 and 3.53\,\mic. Even after a reduction by a factor of three of
the area below the profiles in Figs.\,\ref{fig:3C298a} and
\ref{fig:3C298b}, we estimate that the inferred absence of the bands
would still hold as a significant result, at least for intermediate
values of $\Omega$, between cases A and B.

%ANJA ... and find that it will work if the diamonds doesn't have hydrogen
%bound to the surface, either because they are clean by nature (e.g.
%if they are "larger" and thereby have a smaller surface to volume
%ratio than the smallest diamonds) or because we can argue that they
%will be "cleaned" due to the exposure of the environment (here we
%have to be careful as the limit between cleaning a surface and
%actually destroying the grain might be rather delicate).

{Our conclusion is that hydrogenated nanodiamonds identical to those
found in meteorites are {\it not} present in the mid-IR spectrum of
3C298. An interesting possibility, however, is that the surface
C$-$H bonds have been photo-destroyed by the quasar's intense UV
flux. Such nanodiamonds could not generate the bands discussed in
this paper. Furthermore, the UV properties of the crystallites
should mimic those of cubic terrestrial diamonds rather than those
of meteoritic nanodiamonds (with surface C$-$H bonds).
%This negative result does not rule out the possibility that
%nanodiamonds without surface adsorptions could account for the break.
\citet{haro07}  modify the original B05 dust model by replacing the
meteoritic type by an enlarged size distribution of the cubic
terrestrial type alone. This dust model reproduces the far-UV break
of quasars and would not generate any C$-$H stretch emission. To
verify such a model, we now need to consider bulk impurities of the
C$-$N or C$-$O type \citep{anders99} or the much less efficient
multi-phonon modes \citep{braatz00a, edwards85, JonesHendecourt00}.

\begin{acknowledgements}
This work was funded by the CONACyT grant J-50296 and the UNAM
PAPIIT grant IN118601. It is partly based on observations made with
the {\it Spitzer} Space Telescope, which is operated by the Jet
Propulsion Laboratory, California Institute of Technology under NASA
contract 1407. The Dark Cosmology Centre is funded by the Danish
National Research Foundation. We thank Diethild Starkmeth and
Dominique Binette for help with proofreading.

\end{acknowledgements}

% Acta est fabula -- The show is over

\end{document}